\documentclass[twocolumn,showpacs,preprintnumbers,amsmath,amssymb,nofootinbib,pra,floatfix]{revtex4}
\usepackage{graphicx}
\usepackage{mathrsfs}
\usepackage{subfig}

\newcommand{\half}{\frac{1}{2}}
\newcommand{\eqn}[2][]{\begin{equation}\label{#1}#2\end{equation}}
\newcommand{\ket}[1]{\left|#1\right>}
\newcommand{\bmat}[4]{
\begin{bmatrix}
#1 & #2\\
#3 & #4\\
\end{bmatrix}
}
\newcommand{\paren}[1]{\left(#1\right)}
\newcommand{\cip}[2]{\left<#1|#2\right>}
\newcommand{\coip}[3]{\left<#1\left|#2\right|#3\right>}
\newcommand{\tr}{\text{tr}\,}

\begin{document}

\bibliographystyle{apsrev}

\title{Finite Automata for Caching in Matrix Product Algorithms}

\author{Gregory M. Crosswhite}
\affiliation{Department of Physics\\ University of Washington\\ Seattle, 98185}

\author{Dave Bacon}
\affiliation{Department of Computer Science \& Engineering \\ Department of Physics \\ University of Washington \\ Seattle, 98185}


\email{gcross@phys.washington.edu, dabacon@cs.washington.edu}

\begin{abstract}
A diagram is introduced for visualizing matrix product states which makes transparent a connection between matrix product factorizations of states and operators, and complex weighted finite state automata.  It is then shown how one can proceed in the opposite direction:  writing an automaton that ``generates'' an operator gives one an immediate matrix product factorization of it.  Matrix product factorizations have the advantage of reducing the cost of computing expectation values by facilitating caching of intermediate calculations.  Thus our connection to complex weighted finite state automata yields insight into what allows for efficient caching in matrix product algorithms.  Finally, these techniques are generalized to the case of multiple dimensions.
\end{abstract}

\date{\today}

\maketitle

\newpage

\tableofcontents

\section{Motivation}

Straightforward representations of quantum systems tend to grow exponentially with increasing system size.  Thus, if one wants to have a hope of simulating a large quantum system, one needs to pick a clever method for representing it.  Such a representation needs to have three properties:  it must have a scaling law that makes large systems tractable, it needs to faithfully duplicate the properties of the original system, and it needs to allow one to compute expectation values without returning to the expensive representation.

Matrix product states \cite{Rommer:1997gf,Perotti:2005bh,cond-mat/0404706,cond-mat/0505140,Schollwock:2005ul} have been popular in the last couple of decades because they exhibit all three properties:  they grow linearly with the size of the system, they tend to produce good approximations for many interesting systems \cite{cond-mat/0505140}, and they allow for $O(N)$ calculations of expectations for tensor product operators (where $N$ is the size of the system).  Most operators are not tensor products, but one can always write them as a sum of tensor product operators.  A general operator would require an exponential amount of terms to do this, but fortunately most operators of interest require only $O(N)$ terms.  Thus, the cost of computing an expectation for a matrix product is usually $O(N^2)$.  This result can be improved, however, by using matrix product operators.  Indeed, if one can factor an operator into a matrix product (in addition to factoring the state), then one can reduce the $O(N^2)$ calculation into a $O(N)$ calculation.

In practice, one can do even better than this.  A typical use of a matrix product state is as an ansatz for the variational method\footnote{This idea was originally proposed by Ostlund and Rommer\cite{Ostlund:1995uq}.  It was inspired, however, by the DMRG algorithm (originally proposed by White \cite{White:1992ys}) which has proven to be a very effective means of finding quantum ground states.}.  This technique involves sweeping through the matrices and locally optimizing each site.  Naively, this would require $O(N^2)$ computation time at each site ($N$ terms in the operator, $O(N)$ for each term), but if one performs the ``sweep'' by moving from adjacent site to adjacent site, then one can cache the old results of computations in such a way as to achieve $O(1)$ computation time per site for an overall running time of $O(N)$ per sweep.

In the past, this $O(1)$ behavior has typically been achieved by writing a special caching algorithm for each Hamiltonian \cite{cond-mat/0404706}.  However, by writing a caching code that works with matrix product operators, one can achieve this in a general way for all Hamiltonians;  one need only supply as input a matrix product factorization of an operator.  Thus we see that it is incredibly useful to be able to write down a matrix factorization for operators of interest.  

In this paper, we shall present techniques that help simplify and clarify the construction of matrix product operators.  Our path towards this simplification proceeds by showing that matrix product states and matrix product operators can be thought of as finite complex weighted automata.  This equivalence allows us to recast the problem of supplying a matrix product factorization for an operator as the problem of constructing a complex weighted finite automaton.  The correspondence between matrix product operators and complex weighted finite automata opens up the door for applying techniques across the usually disparate subjects of matrix product algorithms and finite automata.  Thus, for example, operations defined on finite automata, which are regularly used to construct and understand finite automata, can now be applied to matrix product operators.  

The connection we establish between matrix product states and complex weighted finite state automata can be viewed as a formalization of the intuition behind much of the language used to describe these states.  For example, finitely correlated states \cite{fannes:92a}, an early version of matrix product states, were first conceived of by thinking about the tensor index connecting different matrices in matrix product states as a memory used in constructing, one subsystem at a time, a quantum state.  Similarly, much language used to describe matrix product operators speaks of the tensor index connecting different matrices as a signal or correlation between adjacent subsystems.  One of our contributions in this paper is to point out that these intuitions can be formalized in that matrix product states can be thought of as complex weighted finite automata and that this view extends to matrix product operators.  This latter property allows for us to engineer matrix product operators designing finite automata and to apply the techniques and methods of finite automata in this process.

A review of the paper is as follows.  We begin with some background material that presents a pedagogical introduction to matrix product states.  Then we will introduce the key to our method:  a new type of diagram which allows one to visualize matrix product states in a way that makes transparent the type of state that they generate.  Although the application of these diagrams to matrix product states is novel, the diagrams themselves are not:  rather, they will be shown to be merely variants of complex weighted finite state automata.  Once this connection has been made, it will be shown how one can obtain a matrix product factorization of a state or operator by starting with an automaton that generates the ``pattern'' of the operator, and then translating this automaton into a set of matrix factors.  It will then be shown how this process generalizes to multiple dimensions, where the automata connection is particularly insightful.

\section{One-dimensional chains}

\subsection{Background}

Consider a quantum system with $N$ independent observables, such as the $Z$ spin components of a linear one-dimensional chain of spin-$\half$ particles.  In general, the representation of this system must be expressed as a tensor with $N$ indices, $A_{i_1,i_2,\dots,i_N}$.  Each element of this tensor represents the amplitude of a particular system configuration; for example $A_{\downarrow\uparrow\uparrow\downarrow}$ gives the amplitude of a particular system of four particles being in the $\downarrow\uparrow\uparrow\downarrow$ state.  Part of the difficulty in simulating quantum mechanical systems arises from the fact that when one adds another particle to a system, one must add another index to the representing tensor.  Thus, the information needed to represent a quantum state in general grows exponentially with the number of particles.

Fortunately, it turns out that not all quantum states require the full content of an $N$-index tensor.  Some states are special in that they are \textit{separable}, which means that their $N$-index tensor can be factored into the outer-product of $N$ one-index tensors, \eqn[separable]{\Psi_{\alpha\beta\gamma\cdots} = A_\alpha B_\beta C_\gamma \cdots}

This representation is very nice because it grows only linearly with the number of observables;  since it is so nice, in fact, it is not surprising that it comes with a price:  it cannot be used to model systems with any entanglement.

It would be nice to be able to add some entanglement into the above representation in such a way that we do not cause it to revert back to the full $N$-index tensor.  For example, suppose that the observables corresponding to indices $\alpha$ and $\beta$ in \eqref{separable} were maximally entangled -- i.e., their state is given by $$\ket{\uparrow\uparrow}+\ket{\downarrow\downarrow} \equiv
\bmat{1}{0}{0}{1},
$$ where the matrix elements correspond to configuration amplitudes as shown in the following table:

\begin{center}
\begin{tabular}{c||c|c}
               & $\alpha=\downarrow$ & $\alpha=\uparrow$ \\
\hline
\hline
$\beta=\downarrow$ &       1        &       0      \\
\hline
$\beta=\uparrow$   &       0        &       1
\end{tabular}
\end{center}

\noindent (Note that we are not normalizing our states;  this will not be a concern for the purpose of our discussion.)

There is no way to obtain the above matrix by taking the outer-product of two vectors -- $A_\alpha$ and $B_\beta$ in \eqref{separable} -- but one could obtain it by taking the inner product of two matrices, such as $$A_{\alpha\,i} = \bmat{0}{1}{1}{0} = B_{i\,\beta}, \quad \sum_i A_{\alpha i} B_{i \beta} = \bmat{1}{0}{0}{1}. $$  Thus, we see that we could represent our state in the form, $$\Psi_{\alpha\beta\gamma\cdots} = \sum_i A_{\alpha i} B_{i\beta} C_\gamma \cdots$$

This had the desired result -- we were able to add a small amount of entanglement to our separable state without greatly enlarging it.  The inner index $i$ can be thought of as a ``bond'' between two of the particles that allows them to communicate to each other.

If one wished, one could put a bond between all the particles in the (linear one-dimensional) chain, \eqn[mps-oldnotation]{\Psi_{\alpha\beta\gamma\delta\cdots} = \sum_{i,j,k} A_{\alpha i} B_{i \beta j} C_{j \gamma k} D_{k \delta} \cdots,} at which point one would obtain what is called a ``matrix product state''.

This method is not limited to representations of states; it is also possible to likewise factor operators into so-called ``matrix product operators''\cite{Ostlund:1995uq}, $$\Psi_{(\alpha\beta\gamma\delta\cdots)(\alpha'\beta'\gamma'\delta'\cdots)} = \sum_{i,j,k} A_{\alpha\alpha' i} B_{i\beta\beta'j} C_{j\gamma\gamma'k} D_{k \delta\delta'} \cdots.$$

Matrix product states have gained much interest in the last decade because they turn out to have entanglement properties that are sufficient to represent many systems of interest.  Furthermore, they are very flexible:  one can add additional inner indices whenever one wants to introduce entanglement between two particles, forming tensor networks that can represent systems in any number of dimensions and with any lattice structure.

In this paper, it will prove useful to distinguish between two types of indices:  the indices being summed over, which correspond to entanglement introduced between observables, and those not, which correspond to the observables themselves.  Thus, the former will be denoted by subscripts and the latter will be denoted by superscripts; for example, \eqref{mps-oldnotation} should appear in the form, $$\Psi^{\alpha\beta\gamma\delta\cdots} = \sum_{i,j,k} A_i^\alpha B_{ij}^\beta C_{jk}^\gamma D_{k}^\delta \cdots,$$

\subsection{Matrix product diagram}

Consider the four-particle W state $$\ket{\Psi} = \ket{\downarrow\uparrow\uparrow\uparrow} + \ket{\uparrow\downarrow\uparrow\uparrow} + \ket{\uparrow\uparrow\downarrow\uparrow} + \ket{\uparrow\uparrow\uparrow\downarrow},$$ which is a sum over all possible states in which one and only one particle has spin-down.  A matrix product representation of this state is $$\Psi^{\alpha\beta\gamma\delta\cdots} = \sum_{i,j,k} A_i^\alpha B_{ij}^\beta C_{jk}^\gamma D_{k}^\delta,$$ where $\alpha$ is the index of the spin component of the first particle, $\beta$ is the index of the second particle, etc., and the tensors on the right-hand side are given by
$$A^\uparrow = \begin{bmatrix} 1 & 0 \end{bmatrix}, \quad A^\downarrow = \begin{bmatrix} 0 & 1 \end{bmatrix},$$
$$B^\uparrow=C^\uparrow = \bmat{1}{0}{0}{1}, \quad B^\downarrow=C^\downarrow = \bmat{0}{1}{0}{0}.$$
$$D^\uparrow = \begin{bmatrix} 0 \\ 1 \end{bmatrix}, \quad D^\downarrow = \begin{bmatrix} 1 \\ 0 \end{bmatrix},$$

Alternatively, one may use the following notation.  Instead of writing a separate matrix for each value of the superscript indices, instead label each matrix element by a value of the observable.  Furthermore, adopt the convention that when taking the inner-product between matrices, one should multiply the matrix elements together by using the outer-product.  This allows us to express our state in the more compact (and hopefully transparent) form,
\eqn[W-matrix-factorization]{\Psi = \underbrace{\begin{bmatrix}\uparrow & \downarrow \end{bmatrix}}_{A}\cdot
         \underbrace{\bmat{\uparrow}{\downarrow}{0}{\uparrow}}_{B}\cdot
         \underbrace{\bmat{\uparrow}{\downarrow}{0}{\uparrow}}_{C}\cdot
         \underbrace{\begin{bmatrix}\downarrow \\ \uparrow \end{bmatrix}}_{D}}

To illustrate that this factorization of our state works, we step through the multiplication of the matrices, starting with the two on the right:
$$
\begin{aligned}
  \Psi &= \paren{\begin{bmatrix}\uparrow & \downarrow \end{bmatrix}\cdot
          \paren{\bmat{\uparrow}{\downarrow}{0}{\uparrow}\cdot
          \paren{\bmat{\uparrow}{\downarrow}{0}{\uparrow}\cdot
          \begin{bmatrix}\uparrow \\ \downarrow \end{bmatrix}}}}\\
       &= \paren{\begin{bmatrix}\uparrow & \downarrow \end{bmatrix}\cdot
          \paren{\bmat{\uparrow}{\downarrow}{0}{\uparrow}\cdot
          \begin{bmatrix}\uparrow\downarrow+\downarrow\uparrow \\ \uparrow\uparrow \end{bmatrix}}}\\
       &= \paren{\begin{bmatrix}\uparrow & \downarrow \end{bmatrix}\cdot
          \begin{bmatrix}\uparrow\uparrow\downarrow+\uparrow\downarrow\uparrow+ \downarrow\uparrow\uparrow \\ \uparrow\uparrow\uparrow \end{bmatrix}}\\
       &= \uparrow\uparrow\uparrow\downarrow + \uparrow\uparrow\downarrow\uparrow + \uparrow\downarrow\uparrow\uparrow + \downarrow\uparrow\uparrow\uparrow
\end{aligned}
$$

\begin{figure}
\centering
\subfloat[
    Diagram representing the matrix product form of the ``W''-state.  Each possible ``walk'' from left to right generates a term in the state, as illustrated in \ref{fig:examplewalk}
    \label{fig:Wdiagram}
]{\framebox{\includegraphics[width=\columnwidth]{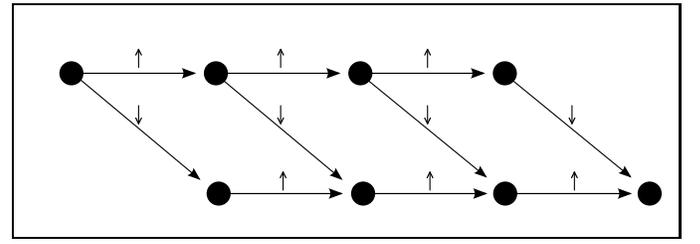}}}
\subfloat[An example walk through the matrix product state illustrated in \ref{fig:Wdiagram}, which generates the term $\uparrow\uparrow\downarrow\uparrow$.
\label{fig:examplewalk}
]{\framebox{\includegraphics[width=\columnwidth]{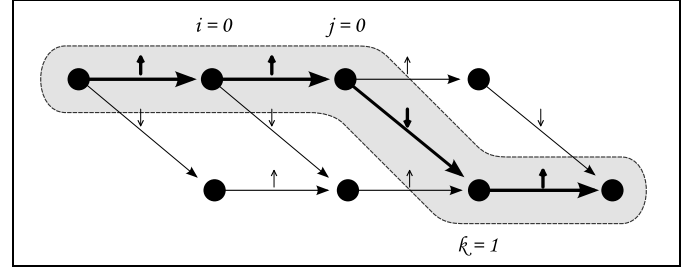}}}
\subfloat[
    Mapping of edges to matrix elements.
    \label{fig:edgemap}
]{\framebox{\includegraphics[width=3in]{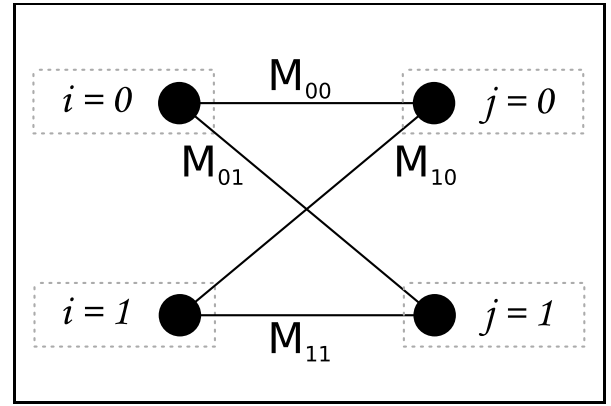}}}
\caption{Matrix product diagrams}
\end{figure}

This factorization may be equivalently expressed in the form of the diagram in figure \ref{fig:Wdiagram}, which was obtained directly from the matrices in \eqref{W-matrix-factorization}.  The nodes correspond to indices, and the edges correspond to matrix elements.  The matrices were treated as a table of weights for edges connecting each set of nodes.  That is, for each $2\times 2$ matrix $M_{ij}$, the elements were mapped to edges as shown in figure \ref{fig:edgemap}.  Where a matrix edge was zero, the edge was omitted.  The nodes shared between edges indicate common indices being summed over.

The arrows place an ordering on the indices.  They are not strictly necessary to define the diagrams, but they are useful because they allow one to view the diagram in terms of paths.  Specifically, each choice of indices corresponds to a ``walk'' from the left side of the diagram to the right.  For example, the choice $i=0, j=0, k=1$ corresponds to the walk as shown in Fig. \ref{fig:examplewalk}.

Each possible walk from the left to the right generates a term in our sum, so that the walk shown in Fig. \ref{fig:examplewalk} generates the term $\uparrow\uparrow\downarrow\uparrow$.  As discussed earlier, edges which do not appear in the diagram correspond to vanishing matrix elements;  this may be thought of as disallowing a walk between certain nodes, as any term which tries to include nonexistent edges is multiplied by zero and thus does not contribute to the sum.  For example, in figure \ref{fig:Wdiagram}, note that there is no path that returns to the top from the bottom (such as $i=0$, $j=1$, $k=0$).

\subsubsection{Extension to operators}

We are not restricted to labeling edges of matrix product diagrams with states;  the tensors at each site may be objects with any number of superscript indices.  This allows us to factor operators as well as states, resulting in a matrix product operator\footnote{Matrix product operators were originally introduced by Verstraete, Garcia-Ripoll and Cirac \cite{cond-mat/0406426}, but they were used as density operators -- i.e., as representations of states, rather than of Hamiltonians or other physical operators.  McCulloch, however, later showed how many classes of physical operators can be factored, and discussed why it can be useful to write them in this form\cite{cond-mat/0701428}.}, which recall has the form
$$\Psi^{(\alpha\beta\gamma\delta\dots)(\alpha'\beta'\gamma'\delta'\dots)} = \sum_{i,j,k,l} A_i^{\alpha\alpha'} B_{ij}^{\beta\beta'} C_{jk}^{\gamma\gamma'} D_{k}^{\delta\delta'} \dots.$$

\begin{figure}
\framebox{\includegraphics[width=\columnwidth]{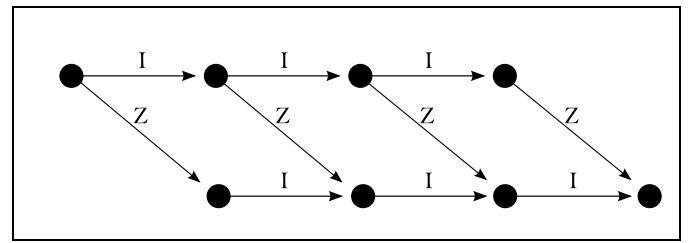}}
\caption{Matrix product diagram for a magnetic field operator. \label{fig:magfield}}
\end{figure}

For example, if we were to take the matrix product representation for the W state, as given in \eqref{W-matrix-factorization}, and replace $\uparrow$ with the identity matrix and $\downarrow$ with the Pauli $Z$ spin matrix (which wewill denote by \textbf{Z}) then we would obtain the diagram in figure \ref{fig:magfield}.

This operator represents the action of a magnetic field in the $z$ direction coupling to each of the particles.

\subsection{Weighted finite automata states}

\label{wfas}

Up to this point we have considered our state to be an $N-$dimensional tensor, where $N$ is the number of observables.  Let us now use a different but equivalent description that is applicable when all of our observables are of the same kind (e.g., $z$ components of spin).  First, we define a set $\Sigma$ to be our ``alphabet'';  it contains all the possible values for our observable.  For example, for a spin-$\half$ chain, we shall choose $\Sigma:=\{0,1\}$ where $0$ labels the spin-up state and $1$ labels the spin-down state.  Then we may describe our state as a function that maps strings of length $N$ of the alphabet $\Sigma$ to complex numbers:  $f: \Sigma^N\to \mathbb{C}$.

We can generalize this function.  Suppose that the size of our system is itself a variable -- that is, we want to consider systems with 1 particle, 2 particles, etc., and to have the descriptions for all of these systems captured in a single function.  Then we can make our function a map not from $\Sigma^N$, but from $\Sigma^*$, the set of all finite-length strings of $\Sigma$ symbols.

When phrased in this form, it can be shown that saying that our state has a matrix product representation is equivalent to saying that the function $f$ can be computed by a special kind of weighted finite automaton.  A complex-weighted finite automaton\footnote{Complex-weighted automata are a generalization of \textit{real}-weighted finite automata, which were originally introduced by Culik and Kari as a technique for compressing grayscale images \cite{II:1993fk,II:fk,Jiang:2003wd}.  It is worth noting that Latorre devised a very similar algorithm for image compression motivated by matrix product states, though without making the connection to finite state automata \cite{quant-ph/0510031}.  This is interesting because it shows how the separate fields of quantum physics and computer science have independently converged to the same idea.} is defined by a 5-tuple, $(Q,\Sigma,W,\alpha,\Omega)$, where

\begin{enumerate}
\item $Q$ is a finite set of \textit{states}
\item $\Sigma$ is a finite \textit{alphabet}; in our case, we shall let $\Sigma=\{0,1\}$ for the two possible values of the $z$ component of spin
\item $W:Q\times\Sigma\times Q\to \mathbb{C}$ is the \textit{weight function};  we may equivalently represent this function as a set of complex $Q\times Q$ matrices, $W_a$ for each symbol $a$ in our alphabet $\Sigma$
\item $\alpha:1\times Q$ is the (complex-valued) \textit{initial distribution}
\item $\Omega:Q\times 1$ is the (complex-valued) \textit{final distribution}
\end{enumerate}

For a string $a_0a_1\dots a_N\in\Sigma^*$, the output of our automaton is defined to be

\eqn[wfa-compute-eqn]{f(a_0a_1\dots a_N) = \alpha \cdot W_{a_0} \cdot W_{a_1} \dots W_{a_N} \cdot \Omega}

A finite state automaton can be thought of as a machine which moves from one one state to another based on an input signal.  To see a simple example of this, we consider a simplification of a weighted finite automata called a \textit{deterministic finite automaton}, which outputs either 0 (``reject'') or 1 (``accept'').  The values of all matrices -- $W_a$, $\alpha$, and $\Omega$ -- are restricted to be either $1$ or $0$, and there may only be one non-zero matrix element of each row of each matrix.  For example, the following machine accepts (or ``recognizes'') all strings that end in either two 0's or two 1's and rejects all others:

\begin{enumerate}
\item $Q:=\{A,B,C,D,E\}$
\item $\Sigma:=\{0,1\}$
\item
$$W_0:=
\begin{bmatrix}
\,\,0\,\,&\,\,1\,\,&\,\,0\,\,&\,\,0\,\,&\,\,0\,\,\\
\,\,0\,\,&\,\,0\,\,&\,\,1\,\,&\,\,0\,\,&\,\,0\,\,\\
\,\,0\,\,&\,\,0\,\,&\,\,1\,\,&\,\,0\,\,&\,\,0\,\,\\
\,\,0\,\,&\,\,1\,\,&\,\,0\,\,&\,\,0\,\,&\,\,0\,\,\\
\,\,0\,\,&\,\,1\,\,&\,\,0\,\,&\,\,0\,\,&\,\,0\,\,\\
\end{bmatrix},\,\, W_1:=
\begin{bmatrix}
\,\,0\,\,&\,\,0\,\,&\,\,0\,\,&\,\,1\,\,&\,\,0\,\,\\
\,\,0\,\,&\,\,0\,\,&\,\,0\,\,&\,\,0\,\,&\,\,1\,\,\\
\,\,0\,\,&\,\,0\,\,&\,\,0\,\,&\,\,0\,\,&\,\,1\,\,\\
\,\,0\,\,&\,\,0\,\,&\,\,0\,\,&\,\,1\,\,&\,\,0\,\,\\
\,\,0\,\,&\,\,0\,\,&\,\,0\,\,&\,\,1\,\,&\,\,0\,\,\\
\end{bmatrix}.$$
\item $\alpha = \begin{bmatrix}\,\,1\,\,&\,\,0\,\,&\,\,0\,\,&\,\,0\,\,&\,\,0\,\,\end{bmatrix}$
\item $\Omega = \begin{bmatrix}0\\0\\1\\0\\1\end{bmatrix}$
\end{enumerate}

\begin{figure}
\framebox{\includegraphics[width=\columnwidth]{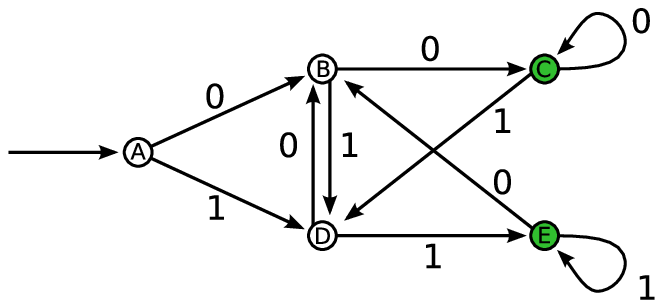}}
\caption{(Color online) Example finite state automaton which recognizes any string that ends in either two 0's or two 1's. \label{fig:automaton1}}
\end{figure}

While this is the canonical form, it is not very transparent.  The diagram in figure \ref{fig:automaton1} is an equivalent method of defining this automaton.  The unconnected arrow on the far left indicates that the system should start in state A.  C and E are shaded to indicate that they are the states that the machine ``accepts'' -- that is, the machine outputs one (as the value for $f$, not to be confused with the symbol 1 in the alphabet) if and only if it is currently in such a state at the end of a string;  otherwise, it outputs zero.  The arrows indicate how the machine should transition between states in response to a symbol;  for example, the machine will move from state A to state B if the first symbol is a 0, and to state D if the first symbol is a 1.

This machine works by using states B through E as a sort of ``memory''.  States B and D are used for the machine to remember that the last symbol it saw (respectively a 0 or a 1) was different from the one before it;  states C and E are used for the machine to remember that it has already seen two symbols of a kind.

Note that for each state there is one and only one transition for each symbol, and this transition has weight 1;  this is due to our restriction that our machine had to be a deterministic finite automaton.  Removal of this restriction allows us to have zero or multiple transitions for each symbol at each state, and also to give a weight to each transition.  Because of this, each input string can have multiple paths, or even no paths through our diagram;  for each possible path we associate a weight equal to the product of all the weights along the path;  furthermore, there may be more than one initial state, and each initial state and final state may itself have a weight.  The output of our automaton is the sum of all weights of all paths from all inital to all final states.  This procedure is not an extension of our definition of a weighted finite automaton, but rather a restatement of it, as it is implicit in \eqref{wfa-compute-eqn}.

Also, note that in this light a matrix product state can be seen as just a special case of a weighted finite automaton.  Each of the nodes on the diagrams drawn earlier is a state, with the edges between them labeling transitions.  In a matrix product state, however, there is a separate transition matrix for each position in the string; that is, the third symbol always passes through the same region on the graph, and no other symbol passes through this same region, whereas in a finite state automaton all states are potentially accessible to all symbols.  (Equivalently, one could say that a matrix product state is a weighted finite automaton in which all the transition matrices are block diagonal.)

\begin{figure}
\subfloat[
    Finite state automaton recognizing the W state.
    \label{fig:Wautomaton}
]{\framebox[\columnwidth][c]{\includegraphics[width=\columnwidth]{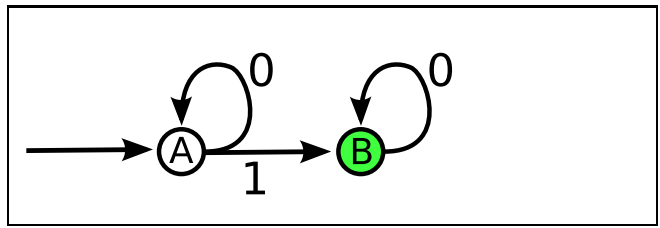}}}
\subfloat[
    Finite state automaton recognizing state with neighboring 1's.
    \label{fig:Neighbor-automaton}
]{\framebox[\columnwidth][c]{\includegraphics[width=\columnwidth]{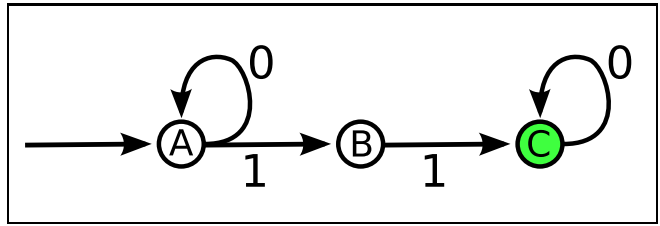}}}
\caption{(Color online)  Examples of finite state automata which recognize quantum states}
\end{figure}

The ability to share states allows one to write down very compact
 representations of states in weighted finite automaton form.  For example, the W state can expressed as an automaton with only two states, as shown in figure \ref{fig:Wautomaton}.

Again, observe that our states act as a form of memory.  When the machine is in state A, it has not yet seen a 1.  When the machine is in state B, it has already seen a 1.  Upon seeing a 1, it either transitions from A to B, or dies if it is already in B (i.e., outputs 0 for the state).  With this manner of thinking, it is easy to see how to extend this machine to output the state $$110000\dots + 011000\dots + 001100\dots + \dots,$$ that is, the set of states with two neighboring 1's;  we already have a state, B, which indicates that the machine has seen one 1, so all we have to do is add another state, C, which indicates that it has seen two 1's.  We also have to update the transitions so that the machine dies unless the two states are neighbors.  The result is shown in figure \ref{fig:Neighbor-automaton}.

\begin{figure*}
\framebox{\includegraphics[width=\textwidth]{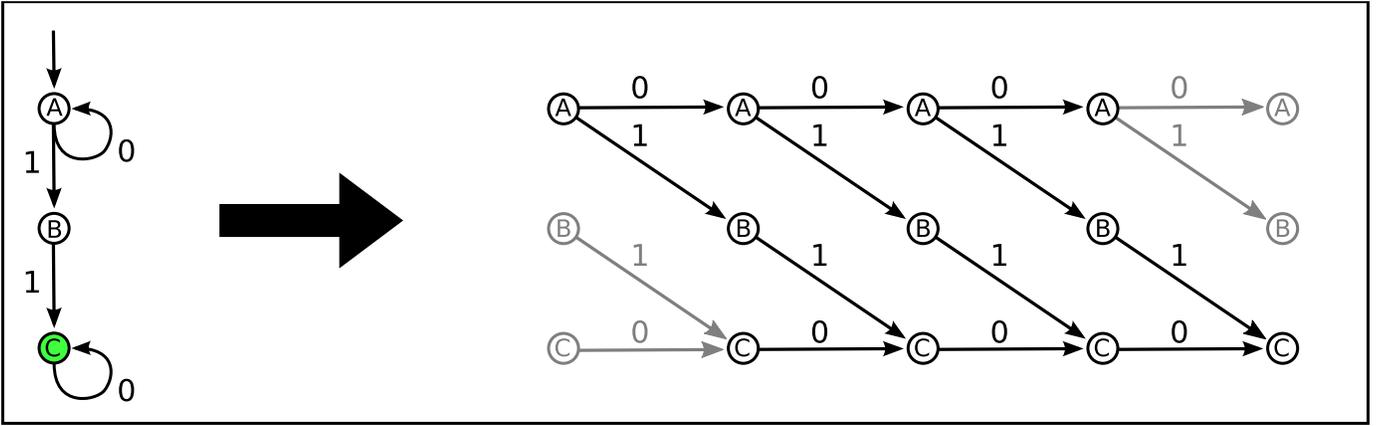}}
\caption{(Color online)  Example of converting a finite state automaton (in this case, for the W-state) into a matrix product state diagram.  Note how some edges have been faded in order to indicate that they have been removed.}
\label{fig:automaton2mps}
\end{figure*}

Just as it is possible to view a matrix product state as a special case of a weighted finite state automaton, it is always possible to construct a matrix product state from a weighted finite automaton.  To do this, one creates a copy of all the states of the automaton for each particle in the system, and remaps the transitions so that one is always moving from one set of states to another.  Finally, one removes all but the initial transition in the leftmost set of states and all but the final transition in the rightmost set of states.  For example, for the state just described the matrix product representation would be as shown in figure \ref{fig:automaton2mps}, where the faded states and edges indicate that they have been removed.

At this point, note that we have obtained a factorization of the nearest neighbor coupling operator, $$XXII + IXXI + IIXX.$$

To see this, we observe that this has the same pattern as the state $$1100 + 0110 + 0011,$$ which is what we just factored above.  Using the same form for the diagram, we see that the matrix factorization is 
$$
\begin{bmatrix}
\textbf{I} & \textbf{X} & 0 \\
\end{bmatrix}
\cdot
\begin{bmatrix}
\textbf{I} & \textbf{X} & 0 \\
0 & 0 & \textbf{X} \\
0 & 0 & \textbf{I} \\
\end{bmatrix}
\cdot
\begin{bmatrix}
\textbf{I} & \textbf{X} & 0 \\
0 & 0 & \textbf{X} \\
0 & 0 & \textbf{I} \\
\end{bmatrix}
\cdot
\begin{bmatrix}
0 \\
\textbf{X} \\
\textbf{I} \\
\end{bmatrix}.
$$

Thus, we see that we now have a method for factoring operators:

\begin{enumerate}
\item Write down a weighted finite automata which generates the pattern of the operator.
\item Translate this into a matrix product operator diagram.
\item Write down matrices based on the diagram.
\end{enumerate}

This method is most efficient for operators that are translationally invariant.  If an operator has additional position-dependent structure, then one should incorporate this structure into the matrix product diagrams, rather than into the weighted finite automata.  To see what is meant by this, suppose our coupling operator took the peculiar form, $$XXII + IX\underline{\textbf{Z}}I + IIXX.$$

\begin{figure}
\subfloat[
    \label{fig:nolongerinvariant-automaton}
]{\framebox{\includegraphics[width=\columnwidth]{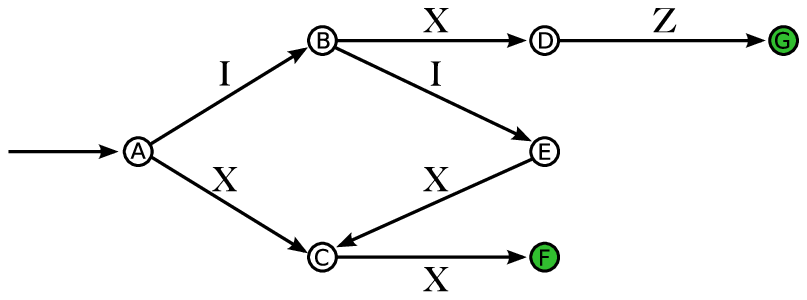}}}
\subfloat[
    \label{fig:nolongerinvariant-mps}
]{\framebox{\includegraphics[width=\columnwidth]{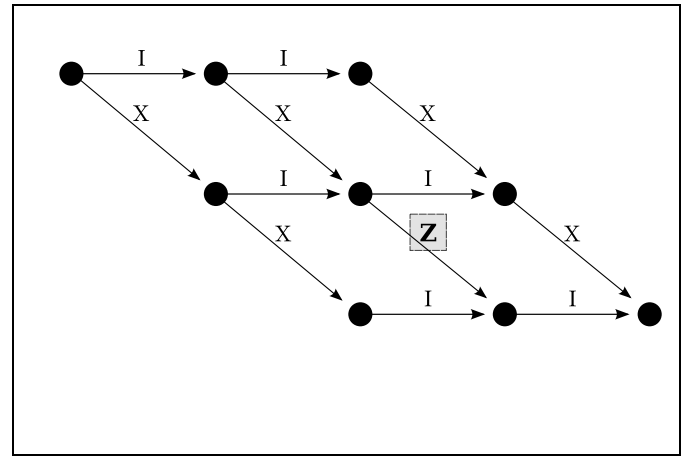}}}
\caption{(Color online)  An example of what happens to an automaton (fig. \ref{fig:nolongerinvariant-automaton}) and a matrix product state (fig. \ref{fig:nolongerinvariant-mps}) when translational invariance is lost.}
\end{figure}

It is still possible to write down a weighted finite automata for this operator, as shown in figure \ref{fig:nolongerinvariant-automaton}.  However, as you can see, capturing this position-dependent behavior requires the addition of several states, which means that our matrix factors would have to be much larger.  Thus, rather than proceeding in this way, it would be better to note that this operator looks almost like the previous operator except with a $Z$ in a special place, and then proceed by modifying the previous diagram in that single spot to obtain the diagram shown in \ref{fig:nolongerinvariant-mps}, which corresponds to the factorization,
$$
\begin{bmatrix}
\textbf{I} & \textbf{X} & 0 \\
\end{bmatrix}
\cdot
\begin{bmatrix}
\textbf{I} & \textbf{X} & 0 \\
0 & 0 & \textbf{X} \\
0 & 0 & \textbf{I} \\
\end{bmatrix}
\cdot
\begin{bmatrix}
\textbf{I} & \textbf{X} & 0 \\
0 & 0 & \fbox{\textbf{Z}} \\
0 & 0 & \textbf{I} \\
\end{bmatrix}
\cdot
\begin{bmatrix}
0 \\
\textbf{X} \\
\textbf{I} \\
\end{bmatrix}.
$$

Thus we see that when encoding translationally invariant behavior in an operator it is better to work with weighted finite automata, and when encoding position dependent behaviour it is better to work with matrix product diagrams.  Of particular significance of this result is that natural operations on finite automata, like constructing unions, concatenations, intersections of their languages, can now be put to use in constructing matrix product operators of increasing complexity.  Recently methods for engineering complex quantum systems, with effective Hamiltonians which are extremely complicated have become important for quantum computation \cite{Kempe:06a,Aharonov:04a,Aharonov:07a}.  Finding matrix product factorizations for these Hamiltonians is a non-trivial task.  However, the above finite automaton picture brings to bear a new set of tools for obtaining such factorizations.

\subsection{Calculation of expectations}

\label{calcexp}

In the proceeding section we have demonstrated a  method for obtaining a factorization of an operator by thinking about these operators as a complex weighted finite automaton.  In this section, we shall show how a matrix factorization of an operator allows us to compute expectations of matrix product states efficiently.

\begin{figure}
\framebox{\includegraphics{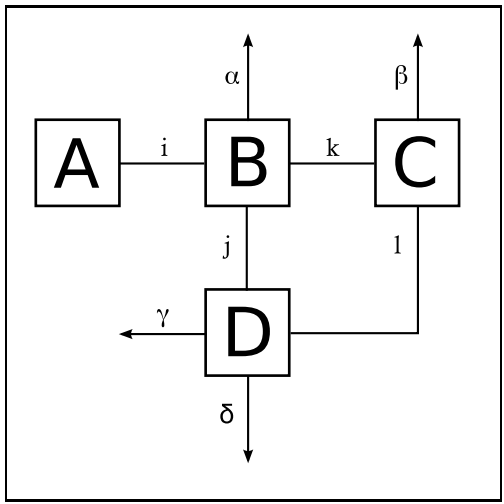}}
\caption{An example of using box-and-line notation to represent a network of tensors.}
\label{fig:box-line-notation-introduced}
\end{figure}

We shall use ``box-and-line'' notation to represent a network of tensors that is being partly or fully contracted, so for example the tensor network $$\Psi^{\alpha\beta\gamma\delta} = \sum_{ijkl} A_{i} B_{ijk}^\alpha C_{kl}^\beta D_{jl}^{\gamma\delta}$$ is represented by the diagram shown in figure \ref{fig:box-line-notation-introduced}, where the boxes represent tensors, edges connecting boxes represent summed (internal) indices, and edges with arrows indicate external indices.

\begin{figure}
\subfloat[
    Matrix product state by itself
    \label{fig:box-line-mps}
]{\framebox{\includegraphics{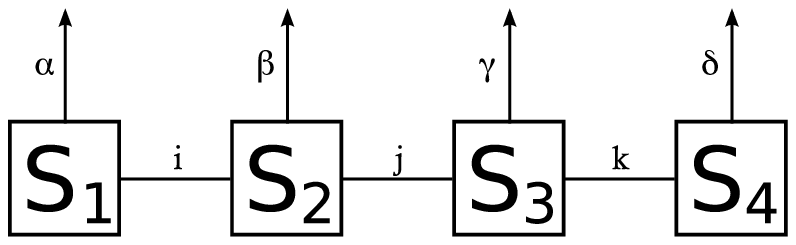}}}
\subfloat[
    Expectation of an arbitrary operator
    \label{fig:box-line-expectation-arbitrary}
]{\framebox{\includegraphics{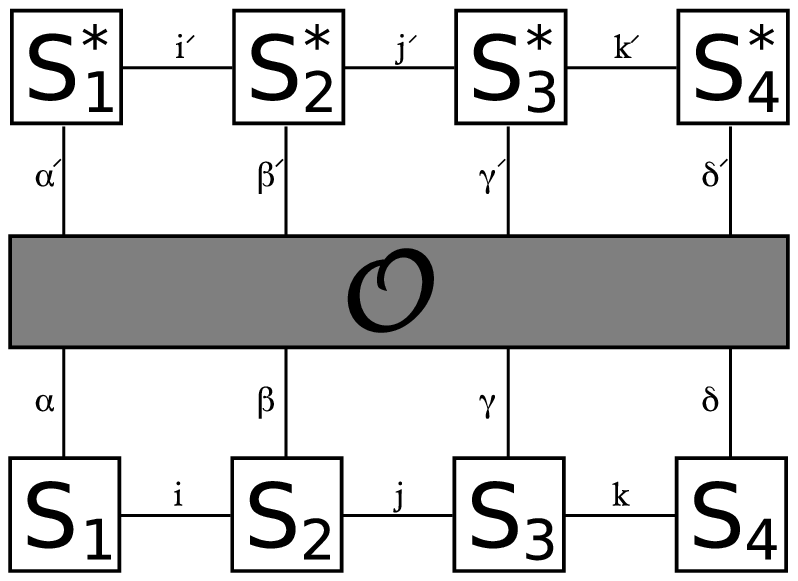}}}
\subfloat[
    Expectation of a matrix product operator.
    \label{fig:box-line-expectation-mpo}
]{\framebox{\includegraphics{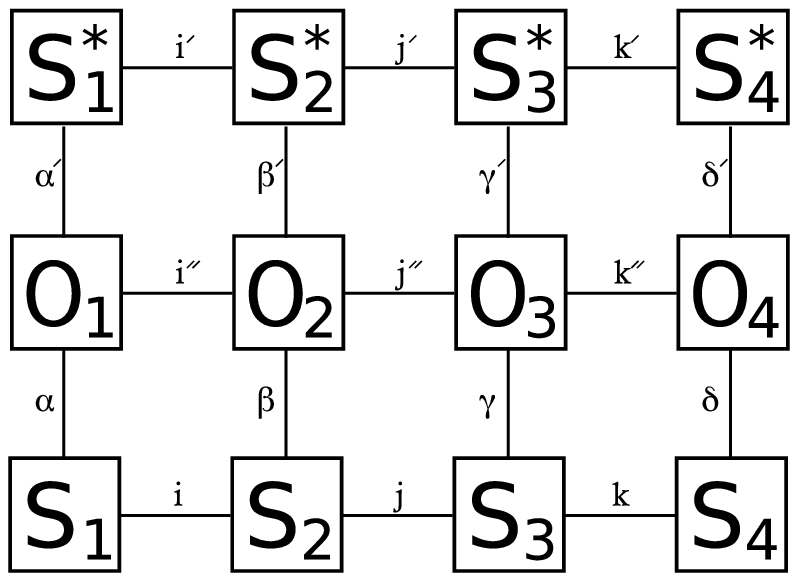}}}
\caption{Box-and-line notation applied to matrix product states.}
\end{figure}

With this notation, we see that the matrix product state given by $$S^{\alpha\beta\gamma\delta} = \sum_{ijk} \paren{S_1}_{i}^\alpha\paren{S_2}_{ij}^\beta\paren{S_3}_{jk}^\gamma\paren{S_4}_{k}^\delta$$ is represented by the diagram in figure \ref{fig:box-line-mps},  and expectation value of this state with respect to some operator $\mathscr{O}^{(\alpha'\beta'\gamma'\delta'),(\alpha\beta\gamma\delta)}$ is given by the diagram in figure \ref{fig:box-line-expectation-arbitrary}.  If this were the best we could do, then the matrix product state would not have helped us very much because we would still need to perform an exponential amount of calculations.  Fortunately, we can improve upon this if we can factor $\mathscr{O}$ into matrix product form,

$$\mathscr{O}^{(\alpha'\beta'\gamma'\delta'),(\alpha\beta\gamma\delta)}
  = \sum_{i'',j'',k''}\paren{O_1}^{\alpha'\alpha}_i\paren{O_2}^{\beta'\beta}_{ij}\paren{O_3}^{\gamma'\gamma}_{jk}\paren{O_4}^{\delta'\delta}_k,
$$

\noindent so that our tensor network now becomes that shown in figure \ref{fig:box-line-expectation-mpo}.

\begin{figure}
\framebox{\includegraphics{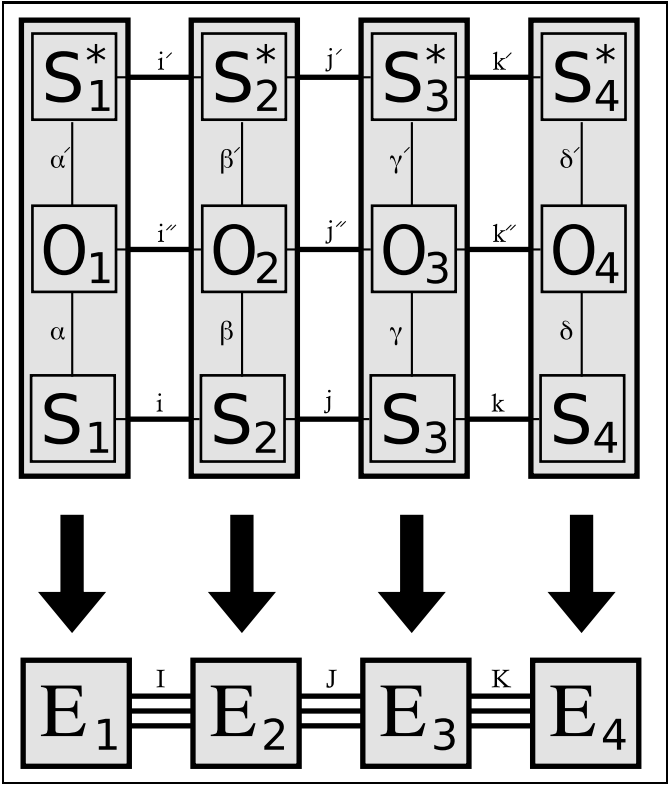}}
\caption{Building transfer matrices for the expectation of a matrix product operator.}
\label{fig:formation-of-transfer-matrices}
\end{figure}

This sum is now performed in two stages;  first, we sum the site and operator matrix at each index to form ``transfer matrices'',
$$\paren{E_1}_I \equiv \paren{E_1}_{(i,i',i'')}
                 = \sum_{\alpha,\alpha'} \paren{S^*_1}^{\alpha'}_{i'}\paren{O_1}^{\alpha'\alpha}_{i''}\paren{S_1}^{\alpha}_i, \dots$$
thus forming the new tensor network shown in figure \ref{fig:formation-of-transfer-matrices}, and then we contract the new network.\footnote{
We note here that there is an alternative viewpoint of matrix product states which considers the transfer matrices themselves to be the primary object of interest.  So-called ``finitely correlated states'' \cite{fannes:92a} are characterized (somewhat abstractly) by a map $\mathbb{E}:\mathcal{A}\otimes\mathcal{B}\to\mathcal{B}$, which in essence produces transfer matrices (tensors in some finite-dimensional tensor space $\mathcal{B}\to\mathcal{B}$) from observables (operators in a $C^*$-algebra $\mathcal{A}$).}
This procedure takes $O(3N)$ matrix multiplications, and the largest matrices that we ever need to form are the $E_n$ matrices.  Thus we see that computing expectation values for a matrix product operator is an $O(N)$ procedure\footnote{Upon completion of this work, we learned of similar results by Murg et al. \cite{0804.3976}.}.

\subsection{Energy minimization and caching}

\label{caching}

Up to now, we have discussed how to perform operations on matrix product states that are known.  In general, however, one will want to investigate systems for which the eigenstates are unknown.  In this case, matrix product states provide an ansatz for the variational method.  That is, one assumes that a ground state has a particular matrix product form, and then searches for the matrix elements which give the lowest energy state representable in that form;  put another way, one seeks the matrix product state $\mathscr{S}$ that minimizes the normalized expectation value of the Hamiltonian
$$f(\mathscr{S}) = \frac{\coip{\mathscr{S}}{H}{\mathscr{S}}}{\cip{\mathscr{S}}{\mathscr{S}}}.$$
The hope is that the result of this process will be a reasonable approximation to the true ground state.

In general finding the global minimizer of $f$ is an NP-complete\footnote{NP-complete problems are the hardest problems in the complexity class NP (nondeterministic polynomial time) and are widely suspected to be computationally intractable.} problem \cite{quant-ph/0609051}.  Fortunately, a local search heuristic suffices for many systems of interest:  at each step in the minimization process, all but one of the site matrices are held constant, and then the energy is minimized with respect to the single remaining matrix.

\begin{figure}
\framebox{\includegraphics[width=\columnwidth]{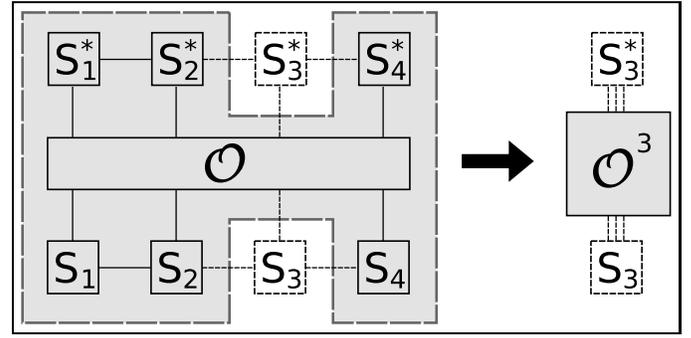}}
\caption{Formation of a matrix that allows us to express the expectation of $\mathscr{O}$ in quadratic form with respect to site 3.}
\label{fig:Omatrix}
\end{figure}

Suppose we are varying over the third matrix.  Recall that the expectation of an operator can be represented by a diagram of the form shown in figure \ref{fig:formation-of-transfer-matrices};  since we are holding all but the third matrix constant, we can form the matrix $\mathscr{O}^{3}$ which is the contraction of all the tensors in the network save $S^3$ and its conjugate, as shown in figure \ref{fig:Omatrix}.  We see now that the energy as a function of $S^3$ is just the quadratic ratio form $$f(S_3) = \frac{S^*_3\cdot \mathscr{H}^3 \cdot S_3}{S^*_3 \cdot \mathscr{N}^3 \cdot S_3},$$ where $\mathscr{H}^3$ and $\mathscr{N}^3$ are the aforementioned tensor contractions for $\mathscr{O}=\mathscr{H}$ and $\mathscr{O}=\textbf{I}$ respectively.  It can be shown that minimizing the above form (a Rayleigh quotient) is equivalent to solving the generalized eigenvalue problem $$\mathscr{H}^3 \cdot S_3 = \lambda \mathscr{N}^3\cdot S_3,$$ and then picking the eigenvector $S^3$ with the smallest value for $\lambda$.  The cost of solving this eigenvalue problem depends only on the size of the site matrix $S^3$, not on the number of sites, $N$;  however, it also relies on $\mathscr{H}^3$, which in general is very expensive to calculate.

\begin{figure}
\framebox{\includegraphics[height=6in]{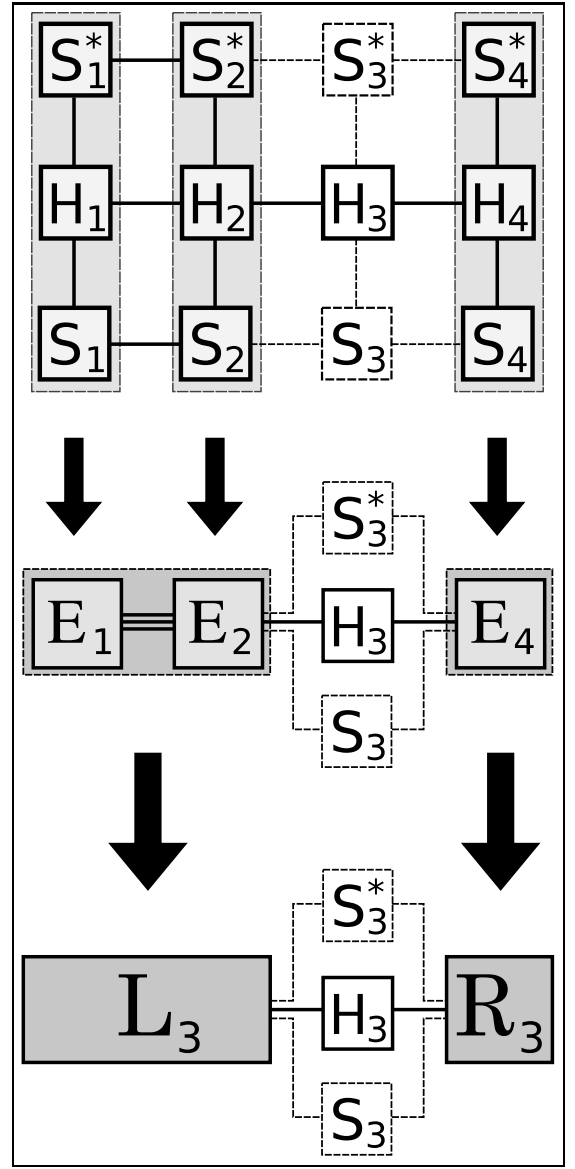}}
\caption{Tensor contractions used to compute $L_3$ and $R_3$}
\label{fig:formation-of-LR}
\end{figure}

Fortunately, if we can factor $\mathscr{H}$ into a matrix product operator, then computing $\mathscr{H}^3$ is cheap.  As in the previous section, we observe that we may form $E_i$ matrices by contracting $S_i^*$, $O_i$, and $S_i$ together at each site;  furthermore, we may also contract all of the $E_i$ matrices to the left of site 3 to form $L_3$ and all of the $E_i$ matrices to the right of site 3 to form $R_3$.  The result of this is the form shown in figure \ref{fig:formation-of-LR}.

Computing $L_i$ and $R_i$ at some site $i$ would naively be an $O(N)$ operation; however, by using caching we can instead make it an amortized\footnote{By ``amortized'' here we mean that although it takes $O(N)$ time to initialize $L_1$ and $R_1$ at the first step, it takes $O(1)$ for all remaining steps, and there are typically at least $N$ steps, so on average the operation takes $O(1)$ time per step.} $O(1)$ operation.  To see why, note that $L_i$ and $R_i$ may be computed recursively:
$$
\begin{aligned}
&L_1 = I,\quad L_{i} = L_{i-1}\cdot E_{i-1},\\
&R_N = I,\quad R_i = R_{i+1}\cdot E_{i+1} \\
\end{aligned}
$$

\begin{figure}
\framebox{\includegraphics[width=\columnwidth]{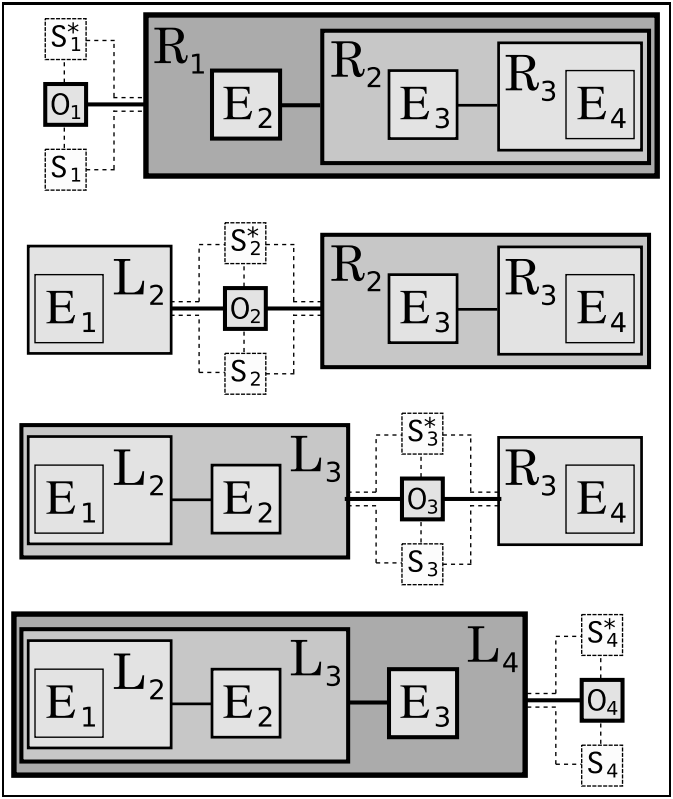}}
\caption{Use of recursion and caching to calculate $L$ and $R$ in amortized $O(1)$ time at each site. \label{fig:recursive-LR} }
\end{figure}

So once we have $R_1$ we already have $R_2$ through $R_N$.  Thus, if we start by minimizing the energy with respect to site 1, and then sweep to the right (i.e., site 2, site 3, up to site N), then although it took us $O(N)$ time to compute $R_1$, we get the $R_i$ matrices for all of the rest of the sites up through $N$ for free.  Once we hit site N, we start moving left back through $N-1,N-2$, etc., and at each step it only takes us one additional matrix multiplication to compute $R_i$ from $R_{i+1}$.  Thus, the time needed at each step to compute $R_i$ is amortized $O(1)$;  by a similar argument, we see the same for the $L_i$ matrices.  This process is illustrated in figure \ref{fig:recursive-LR}.

The notion of using caching to speed up these calculations is not a new one;  the same process has already been described by Verstraete, Porras, and Cirac \cite{cond-mat/0404706}.  However, whereas their process is limited to one- and two-body operators, our procedure works for any form of operator that can be written in matrix product form;  furthermore, the process is the same for all such operators, rather than requiring a new process for each class of operator -- e.g., one-body, two-body, etc.

To summarize: if we can factor our Hamiltonian into a matrix product operator, then we can calculate our matrices $\mathscr{H}^i$ and $\mathscr{N}^i$ needed to optimize a site matrix from matrices that can be calculated using a recursion rule.  By caching the intermediate steps of the recursion, and moving through the sites to be optimized in order from left to right and back, we can calculate $\mathscr{H}^i$ and $\mathscr{N}^i$ in amortized $O(1)$ time.

\section{Arbitrary-dimensional graph}

\subsection{Motivation}

\label{arbitrary-motivation}

\begin{figure}
\subfloat[
    Total ordering imposed on two-dimensional grid
    \label{fig:2d-total-ordering}
]{\framebox[2.5in]{\includegraphics[width=2.25in]{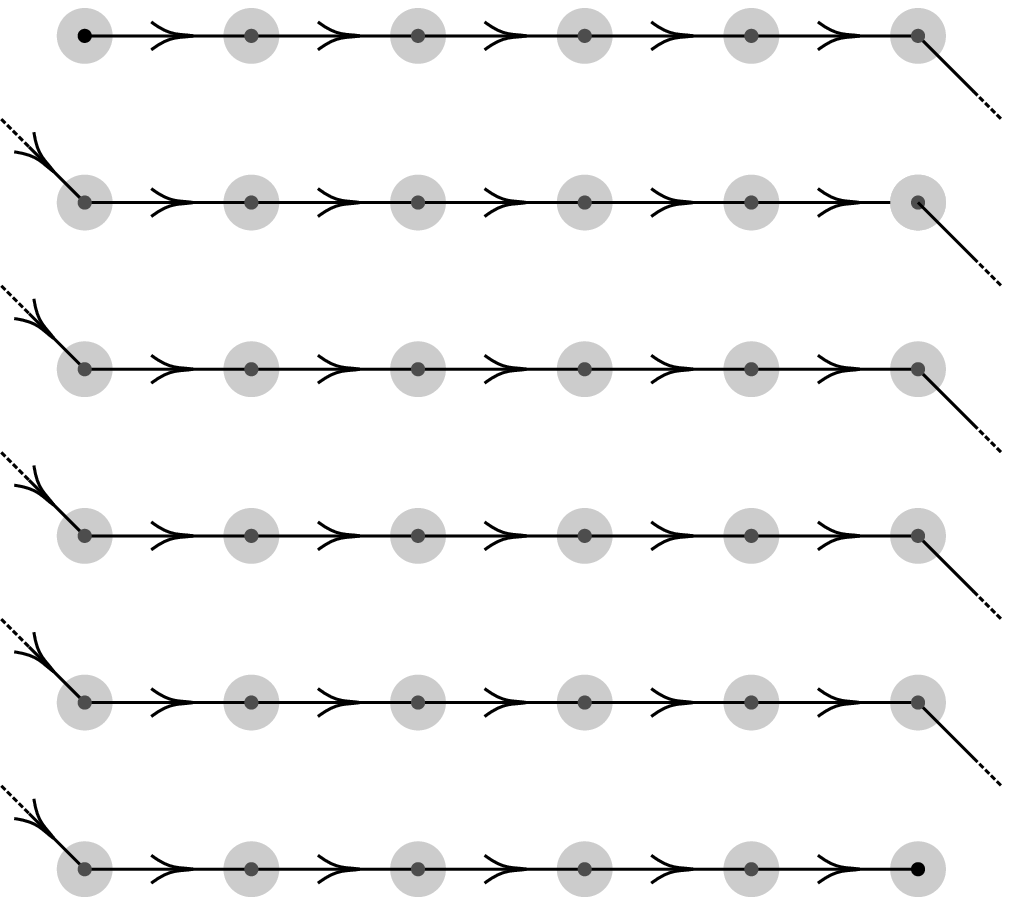}}}
\subfloat[
    4-$X$ operator on grid.  The $X$'s represent the locations of the $X$ operators on the grid; at all other sites there are $I$ operators.
    \label{fig:2d-4X-operator}
]{\framebox[2.5in]{\includegraphics[width=2in]{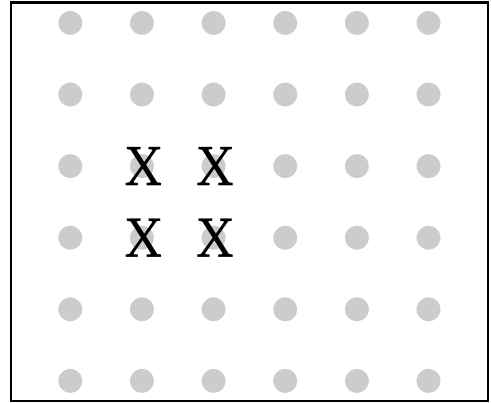}}}
\vspace{.5in}
\subfloat[
    Finite state automaton needed for operator in \ref{fig:2d-4X-operator} given the total ordering shown in \ref{fig:2d-total-ordering}
    \label{fig:2d-4X-FSA}
]{\framebox{\includegraphics[height=3in]{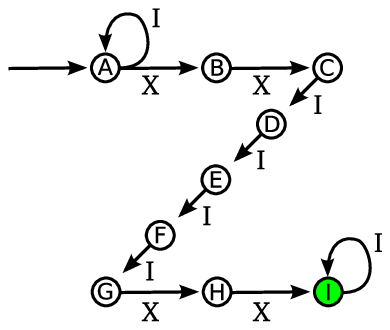}}}
\caption{(Color online)  A total ordering is imposed on the two-dimensional grid in \ref{fig:2d-total-ordering};  this allows us to write down a finite state automaton representation of the 4-$X$ operator shown in \ref{fig:2d-4X-operator}, but the resulting automaton in \ref{fig:2d-4X-FSA} is less than ideal.}
\end{figure}

Matrix product states were designed for studying one-dimensional systems, and that is where they excel.  Nothing technically stops one from using them to study higher-dimensional systems, however -- as long as one is willing to effectively reduce these systems into a one-dimensional system by imposing an ordering; for example, on a $6\times 6$ two-dimensional grid one could impose the ordering shown in figure \ref{fig:2d-total-ordering}.

However, there is a price one pays for doing this.  Suppose that one wants to represent a Hamiltonian on a $6\times 6$ grid which consists of four-site $X$ terms arranged in a square as shown in Fig. \ref{fig:2d-4X-operator}. The automata which encodes such a Hamiltonian takes the form shown in Fig. \ref{fig:2d-4X-FSA}.  The states in the middle (D-G) act as a memory which tells the automata how many sites it has walked past since the second $X$.  This is needed so that the automaton can put the last two $X$'s in the correct place on the following row.  The number of states required here grows with the number of columns in the grid.

We see that although we can write down such an automata, and so form a matrix product representation of this operator, it is less than ideal because the representation depends on the size of the grid.  This comes from the fact that information can only flow in one direction;  ideally, we would like the information that an $X$ has been placed on one row to somehow go directly down one row rather than having to sweep through the rest of the current row first.  We could, of course, adjust the ordering to sweep down columns instead of across rows, but then we lose the ability to cheaply send information across a row;  using matrix product states, there is no way we can make it easy to communicate in two directions simultaneously.

\subsection{Tensor network diagrams}

\label{tensordiagram}

The previous section has described the limitations of matrix product states.  These limitations come from the fact that each tensor is connected by indices to only two other tensors.  (Or equivalently, we might say that the problem is that each site is directly entangled with only two other sites.)  We can get a more powerful representational form, as for example was done using the concept of projected entangled pairs in \cite{cond-mat/0407066,quant-ph/0601075}, by using a more complicated index structure; for example, we could use the following structure:
\eqn[crazy-tensor-product]{\Psi^{\alpha\beta\gamma\delta\mu\nu} =
  \sum_{ijkl} A^\alpha B^{\beta}_{i} C^{\gamma}_{ijk} D_{l} E_{jl}^{\delta\mu} F_{kl}^{\nu}.}

In this example, we see that there are many different types of factors that are possible.  The first, $A^\alpha$, is a simple outer-product factor;  this indicates that there is no entanglement between $\alpha$ and any of the other observables.  The second two tensors, $B^\beta_i$ and $C^\gamma_{ij}$, are connected by an inner product -- i.e., a sum over the subscript index $i$ -- and so we see that $\beta$ and $\gamma$ have some entanglement between them.  $\gamma$ is also entangled with $\delta$ and $\mu$ through the index $j$ and $\nu$ through the index $k$;  this illustrates that entanglement may be shared between one observable and any number of others, and that those other observables need not be directly adjacent to a factor in the above.  Note that the factor $D$ does not have a superscript;  it does not give any direct information about an observable, but rather (in a manner of speaking) it coordinates communication between observables.  On the other hand, $E_{jl}^{\delta\mu}$ has two superscripts, so that it gives information about two observables at once.  Finally, note that the index $l$ is shared between three tensors;  putting the same index in multiple places allows several observables to be simultaneously entangled with each other.

\begin{figure}
\subfloat[
    A diagram representing the tensor product structure of eqn. \eqref{crazy-tensor-product}.
    \label{crazy-diagram}
]{\framebox{\includegraphics[width=\columnwidth]{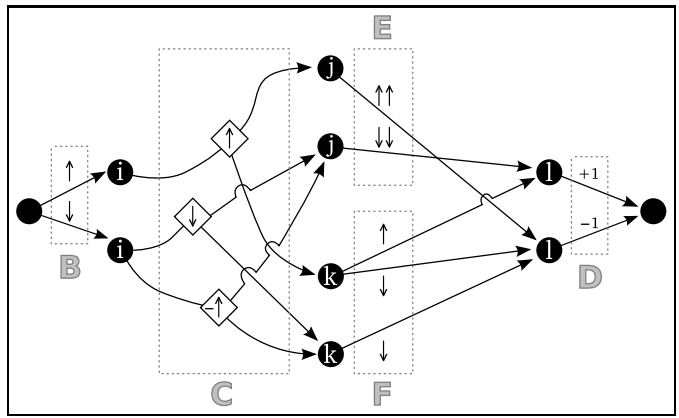}}}
\subfloat[
    An invalid walk through \ref{crazy-diagram}.
    \label{crazy-diagram-bad-walk}
]{\framebox{\includegraphics[width=\columnwidth]{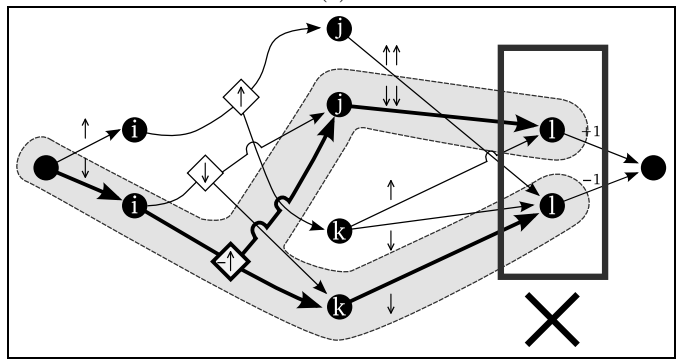}}}
\subfloat[
    An acceptable walk through \ref{crazy-diagram}.
    \label{crazy-diagram-good-walk}
]{\framebox{\includegraphics[width=\columnwidth]{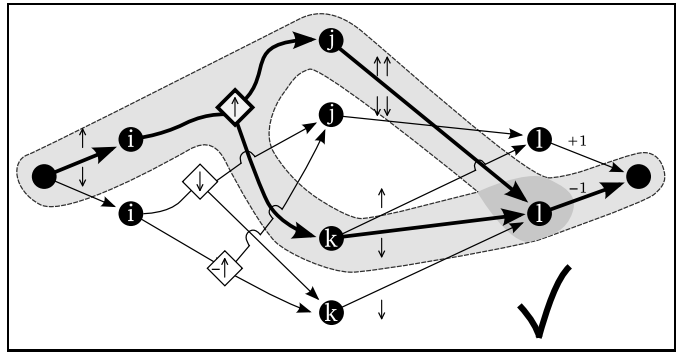}}}
\caption{Tensor product state diagrams}
\end{figure}

One possible diagram with the above tensor structure is that shown in Fig. \ref{crazy-diagram}.  Just as in the one-dimensional matrix product states, we may put arrowheads on the edges, and think of our states as being generated by walks through the diagram.  However, now there are points where our walk may split into many paths (in this case, the indices $j$ and $k$) which are taken simultaneously.  Whenever these two paths rejoin, they must rejoin at the same node or else the walk is rejected.  For example, Fig. \ref{crazy-diagram-bad-walk} illustrates an invalid choice of path, whereas Fig. \ref{crazy-diagram-good-walk} illustrates a correct choice.

This rule for the rejoining of paths is just a restatement of the fact that only one value may be picked for each index for each term in the sum.

Note that although there is a partial ordering on the steps in our walk, there is not a total ordering.  That is, although our arrowheads tell us that a link from C must be chosen before a link from E or F, they do not tell us whether a link from E should be chosen before F or vice versa.  This contrasts with the one-dimensional case, where there is a total ordering.

\subsection{Weighted finite signaling agents}

Recall that in \ref{wfas}, we showed that matrix product diagrams are equivalent to defining a weighted finite automata which encodes the state.  If we wanted, we could similarly relate our generalized tensor network states to weighted finite automata.  There is a catch, though:  automata require the system to be in a concrete ``state'' at any moment in time, and they also require there to be a total ordering of the input.  Our tensor network state diagrams have neither of these properties.

To see why these properties are absent, we return to figure \ref{crazy-diagram}.  For the B transition, the system picks one of the states in $i$ in response to the first input symbol, so both properties hold.  For the C transition, however, the system picks two new states for the system -- one from $j$ and one from $k$ -- in response to the second input symbol.  At this point, not only is the system in multiple states, but the symbol to which it will respond next is not defined, since the order of E and F has not been specified.

Of course, we could force these properties to be present by combining indices $j$ and $k$ into a single index $m$ that unifies them.  (i.e., $m=1$ would be equivalent to $j=1, k=1$;  $m=2$ would be equivalent to $j=1,k=2$, etc.)  We would then have to replace our tensors $E$ and $F$ with a single tensor $G$.  It is possible that in this particular situation we would obtain something simpler, but in general this process will result in much larger and more complicated tensors than we started with.

Thus, instead of thinking in terms of states, it proves more useful to think in terms of signals.  Each site corresponds to an agent that receives signals from channels, makes a (nondeterministic, weighted) decision based on these incoming signals and an input symbol, and then sends signals to output channels.  $i$, $j$, $k$, $l$, and the rest of the unlabeled nodes above correspond to our channels, and $B-F$ correspond to our agents.

\begin{figure}
\framebox{\includegraphics[width=3in]{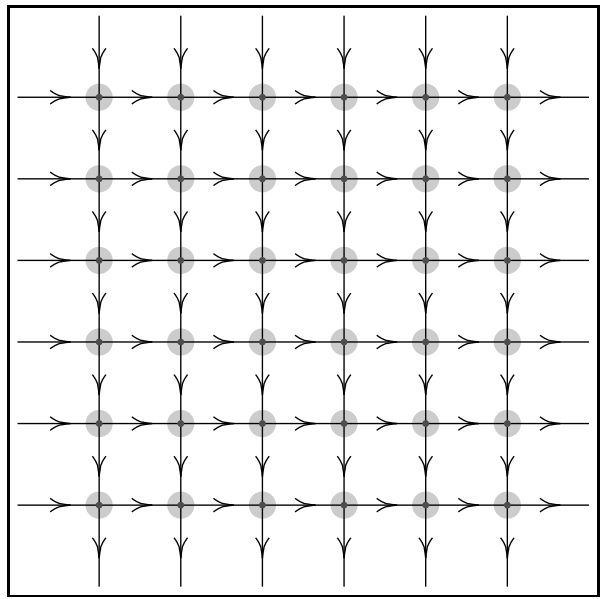}}
\caption{Flow of information for finite signaling agent on two-dimensional grid}
\label{fig:2d-information-flow}
\end{figure}

To see how this works in practice, we consider how one might design an agent to generate the 4-$X$ Hamiltonian (fig. \ref{fig:2d-4X-operator}) discussed in section \ref{arbitrary-motivation}.  We allow our agents to receive signals from two directions: up and left, and to send signals in two directions:  down and right;  the flow of information is illustrated in figure \ref{fig:2d-information-flow}.

The grey circles represent sites (or ``agents'') in our system, and the arrows represent links (or ``channels'').  At each site is an agent, which is a rank 5 tensor with an index corresponding to each channel and an index corresponding to the input symbol.  The (slightly grayed) arrows on the outside of the diagram that connect to only one node implement the boundary conditions;  they do this by starting the system with a particular set of ``initial'' signals, sent through the top and left boundary channels, and then accepting only those inputs that cause the signal received from the bottom and right boundary channels to be one of the valid ``final'' signals.

Each signal is an integer that corresponds to an index in a tensor;  it often proves convenient, though, to map these integers to names in order to make clear the working of the agent.  For example, table \ref{tbl:assign} gives the signal names that we use for the agent recognizing our 4-$X$ Hamiltonian.

\begin{table}

\subfloat[For convenience, labels are assigned to index numbers in order to give intuitive names to the signals.
    \label{tbl:assign}
]{
\begin{tabular}{l|l}
Name    & Index Number \\
\hline
Exterior & 0 \\
Boundary w/ $X$ & 1 \\
Boundary & 2 \\
Interior w/ $X$ & 3 \\
Interior & 4
\end{tabular}}

\subfloat[The transition table defining the finite signaling agent.
    \label{tbl:transitions}
]{
\begin{tabular*}{\columnwidth}{c@{\hspace{.1in}} lcll}
& Input Signals & Symbol & Output Signals \\
\hline
\hline
 & \parbox{0.9in}{\noindent
\begin{trivlist}
\item[$\uparrow$] Exterior
\item[$\leftarrow$] Exterior
\end{trivlist}}
 & $\longmapsto \,\, I \,\, \longmapsto$ &
\parbox{0.9in}{\noindent
\begin{trivlist}
\item[] Exterior $\rightarrow$
\item[] Exterior $\downarrow$
\end{trivlist}}
\\
\hline
\parbox{1cm}{\includegraphics{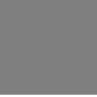}} & \parbox{0.9in}{
\begin{trivlist}
\item[$\uparrow$] Exterior
\item[$\leftarrow$] Exterior
\end{trivlist}}
 & $\longmapsto \,\, X \,\, \longmapsto$ &
\parbox{1.3in}{\noindent
\begin{trivlist}
\item[] Boundary w/ $X$ $\rightarrow$
\item[] Boundary w/ $X$ $\downarrow$
\end{trivlist}}\\
\hline
\parbox{1cm}{\includegraphics{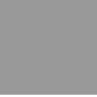}} & \parbox{1.2in}{
\begin{trivlist}
\item[$\uparrow$] Boundary w/ $X$
\item[$\leftarrow$] Exterior
\end{trivlist}}
 & $\longmapsto \,\, X \,\, \longmapsto$ &
\parbox{1.2in}{\noindent
\begin{trivlist}
\item[] Interior w/ $X$ $\rightarrow$
\item[] Boundary $\downarrow$
\end{trivlist}}\\
\hline
\parbox{1cm}{\includegraphics{grey-40}} &  \parbox{1.25in}{
\begin{trivlist}
\item[$\uparrow$] Exterior
\item[$\leftarrow$] Boundary w/ $X$
\end{trivlist}}
 & $\longmapsto \,\, X \,\, \longmapsto$ &
\parbox{1.2in}{\noindent
\begin{trivlist}
\item[] Boundary $\rightarrow$
\item[] Interior w/ $X$ $\downarrow$
\end{trivlist}}\\
\hline
\parbox{1cm}{\includegraphics{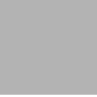}} & \parbox{0.9in}{
\begin{trivlist}
\item[$\uparrow$] Exterior
\item[$\leftarrow$] Boundary
\end{trivlist}}
 & $\longmapsto \,\, I \,\, \longmapsto$ &
\parbox{1in}{\noindent
\begin{trivlist}
\item[] Boundary $\rightarrow$
\item[] Interior $\downarrow$
\end{trivlist}}\\
\hline
\parbox{1cm}{\includegraphics{grey-30}} & \parbox{0.9in}{
\begin{trivlist}
\item[$\uparrow$] Boundary
\item[$\leftarrow$] Exterior
\end{trivlist}}
 & $\longmapsto \,\, I \,\, \longmapsto$ &
\parbox{1in}{\noindent
\begin{trivlist}
\item[] Interior $\rightarrow$
\item[] Boundary $\downarrow$
\end{trivlist}}\\
\hline
\parbox{1cm}{\includegraphics{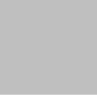}} & \parbox{1.1in}{
\begin{trivlist}
\item[$\uparrow$] Interior w/ $X$
\item[$\leftarrow$] Interior w/ $X$
\end{trivlist}}
 & $\longmapsto \,\, X \,\, \longmapsto$ &
\parbox{1in}{\noindent
\begin{trivlist}
\item[] Interior $\rightarrow$
\item[] Interior $\downarrow$
\end{trivlist}}\\
\hline
\parbox{1cm}{\includegraphics{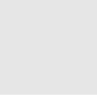}} & \parbox{0.9in}{
\begin{trivlist}
\item[$\uparrow$] Interior
\item[$\leftarrow$] Interior
\end{trivlist}}
 & $\longmapsto \,\, I \,\, \longmapsto$ &
\parbox{1in}{\noindent
\begin{trivlist}
\item[] Interior $\rightarrow$
\item[] Interior $\downarrow$
\end{trivlist}}\\
\end{tabular*}
}

\caption{These tables define a finite signaling agent which recognizes the 4-$X$ Hamiltonian.}

\end{table}

We define an agent by specifying how it reacts to incoming signals.  In this case, there are two incoming signals:  one from above, and one from the left.  Since the agent is nondeterministic, it can have several possible reactions to the incoming signals, each corresponding to a symbol it recognizes (or generates) and signals that it sends right and down.  For our 4-$X$ Hamiltonian, our agent takes the form defined in table \ref{tbl:transitions}.

\begin{figure}
\subfloat [
    A term accepted by the agent
    \label{fig:active-agent-good}

]{\framebox{\includegraphics[height=3in]{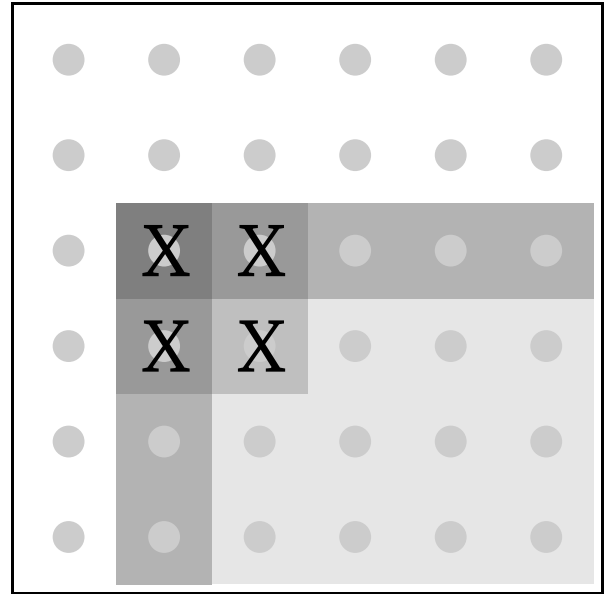}}}
\subfloat [
    A term rejected by the agent due to intersecting boundaries
    \label{fig:active-agent-bad}
] {\framebox{\includegraphics[height=3in]{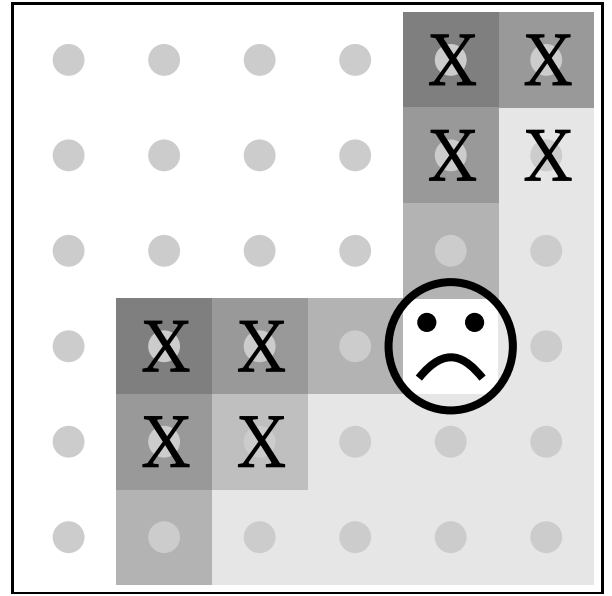}}}
\caption{Transitions experienced by an agent for two possible terms}
\end{figure}

To see what is going on, consider figure \ref{fig:active-agent-good}, which illustrates the agent accepting four $X$'s on the grid.  The background at each point is shaded to indicate which of the above transitions is taking place at that point.  Note that the grid is divided into three general regions:  the exterior (shaded white), the boundary (shaded dark grey), and the interior (shaded light grey).  Inside the boundary and the interior, $X$s are excluded because there is no transition that includes them.  The boundary has the role of forbidding additional squares of $X$s from being accepted in the exterior, since we have chosen our transitions so that boundaries can only be continued in one direction (vertical or horizontal).  As figure \ref{fig:active-agent-bad} illustrates, a second group of $X$s which is in the exterior of the first group results in an intersecting boundary that causes the pattern to be rejected.

Now that we have written down an agent that accepts squares with 4 $X$s, we see that we have immediately obtained a factorization of the Hamiltonian which contains a term for each possible placement of these operators.  This factorization---which can be thought of as a ``projected entangled-pair operator'' (i.e., the natural generalization of projected entangled-pair states to represent operators)---is a tensor network with the tensors located at the grid points;  links between nodes indicate that the corresponding indices of the two tensors should be summed over.  The tensors at each node are of rank 6 -- four of the dimensions correspond to the links, and two correspond to the physical quantum operator.  There are only eight non-zero elements of this tensor, corresponding to the eight entries in Table \ref{tbl:transitions}: $(0,0,0,0,I), (0,0,1,1,X), (1,0,3,2,X),$ etc.

\subsection{Calculation of expectations using recursion}

Assume that we have a tensor network state and a factorization of an operator which has the same network structure as the state.  As in section \ref{calcexp}, we see that the expectation of the operator may be reduced to the contraction of a network of ``transfer matrices'';  for example, for the peculiar state shown in section \ref{tensordiagram}, calculating the expectation of our operator is equivalent to contracting a tensor network of the form shown in figure \ref{crazy-diagram}.\footnote{Again, there is an alternative viewpoint that considers the transfer matrices themselves to be the primary object of interest.  In particular, there is a generalization of finitely correlated states, known as ``contour correlated states'' \cite{Richter:1994hc}, which characterizes quantum states on a two-dimensional lattice by a map $\mathbb{E}:\mathcal{A}^\Lambda\otimes \mathcal{B}^{\partial_{-}\Lambda}\to \mathcal{B}^{\partial_{+}\Lambda}$ that can be thought of as taking observables in the lattice region $\Lambda$ (members of a $C^*$-algebra $\mathcal{A}^\Lambda$) to a transfer matrix in the tensor space $\mathcal{B}^{\partial_{-}\Lambda}\to \mathcal{B}^{\partial_{+}\Lambda}$ which transports information through the region $\Lambda$ from its lower-left contour $\partial_{-}\Lambda$ to its upper-right contour $\partial_{+}\Lambda$.  This lattice is dual to the lattice used by our agents, so the case where the region $\Lambda$ is a single tile corresponds to a transfer matrix derived from an agent on the corresponding vertex in our lattice, with the minor difference that our agents use a different direction (upper-left to lower-right) for the flow of information on the lattice.}

\begin{figure}
\centering
\framebox{\includegraphics[width=\columnwidth]{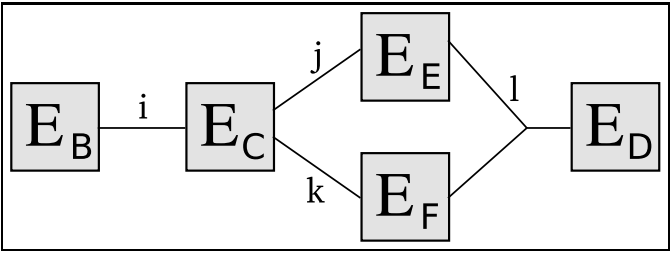}}
\label{crazy-transfer-matrix-network}
\caption{Tensor network giving the expectation of some operator with respect to the site shown in figure \ref{crazy-diagram}.  Note that the index $l$ connects three tensors.}
\end{figure}

We may wish to minimize the energy with respect to some site $n$.  As discussed in section \ref{caching}, we can reduce this to an eigenvalue problem for the matrix consisting of the contraction of (essentially) all of the transfer matrices except the one at $n$.  This contraction can be expressed as a set of recursion rules;  for example, for a two-dimensional grid we have the following rules:
$$
\begin{aligned}
O_{i,j} &= L_{i,j} \cdot A_{i,j} \cdot B_{i,j} \cdot R_{i,j}, \\
L_{i,j} &= L_{i-1,j}\cdot C_{i-1,j}, \quad L_{1,j} = I \\
R_{i,j} &= R_{i+1,j}\cdot C_{i+1,j}, \quad R_{N,j} = I \\
C_{i,j} &= A_{i,j}\cdot E_{i,j} \cdot B_{i,j} \\
A_{i,j} &= A_{i,j-1}\cdot E_{i,j-1}, \quad A_{i,1} = I \\
B_{i,j} &= B_{i,j+1}\cdot E_{i,j+1}, \quad B_{i,N} = I \\
\end{aligned}
$$
(The $\cdot$ operation is implicitly over only the connected indices.)

\begin{figure}
\centering
\includegraphics[width=\columnwidth]{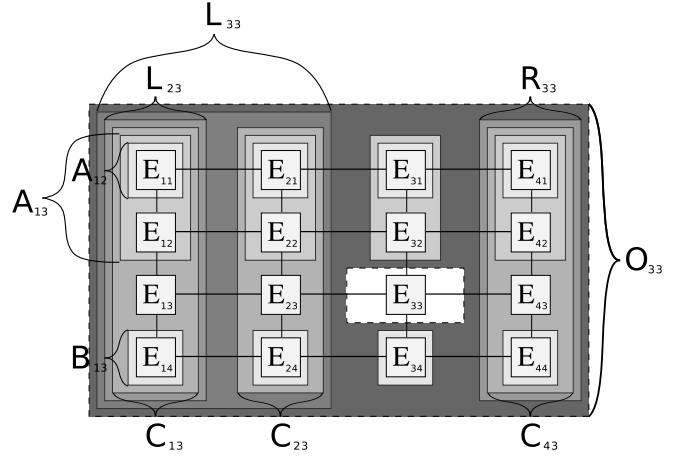}
\caption{Use of recursion to calculate $\mathscr{O}_{33}$. \label{fig:2d-recursive-structure}}
\end{figure}

The computation of $O_{33}$ is illustrated in figure \ref{fig:2d-recursive-structure}.

As was the case in section \ref{caching}, as long as we move from each site to an adjacent site, it takes us only (amortized) $O(1)$ time to calculate $\mathscr{O}_{ij}$.  In Fig. \ref{fig:2d-recursive-structure}, for example, we see that to compute $\mathscr{O}_{43}$, we need only calculate $C_{33}$ and then $L_{43}$.

Unfortunately, it is intractable to contract arbitrarily large multi-dimensional tensor networks.  (Formally, Schuch et al. \cite{quant-ph/0611050} have shown that this is a \#P-complete\footnote{\#P-complete problems are the counting equivalent of NP-complete problems, and are widely thought to be computationally intractable.} problem.)  This is because whenever one contracts together tensors with more than two indices, one obtains a larger tensor.  For example, when taking the dot-product between two four-index tensors one obtains a six-index tensor, $$\sum_f A_{abcd} B_{defg} = C_{abcefg}.$$  These extra indices result in ``double-bonds'' between tensors.  (We have already seen multiple bonds when computing the $E$ matrices, as shown in figure \ref{fig:formation-of-LR}.)

Thus, as we contract each row, the size of our tensors increases by some factor, which means that the cost of contracting a tensor network in general grows exponentially with the size of the network!  Fortunately, there is a lossy compression technique which involves approximating a row resulting from a contraction with a new row with fewer bonds;  this has been used successfully to model hard-core bosons in a two-dimensional optical lattice\cite{cond-mat/0611522}.

\section{Conclusion}

In this paper, we have introduced a type of diagram for representing matrix product states.  We used this to demonstrate that there is a formal equivalence between matrix product states and operators and complex-weighted finite state automata.  This equivalence was used to present a method by which one could factor a matrix product operator by reasoning about these complex weighted finite state automata.  We then showed how such a matrix product factorization of an operator allows one to compute expectations of that operator in $O(N)$ time, and to perform energy minimization at an amortized cost of only $O(1)$ per step.  A generalization of this procedure was presented that allows one to carry the same process through for systems with more than one spatial dimensions.

As a closing remark, we note that this formalism is interesting not only because of its practical application in simulating physical systems, but also because it relates our ability to efficiently simulate physical systems with a broader theory (the automata hierarchy or formal language theory) which deals with the fundamental limits of computation.  It would be interesting to see whether there are other insights from this theory that could be used to improve techniques for simulating physical systems.


\acknowledgements{
Gregory Crosswhite and Dave Bacon were supported under NSF Grant No. 0523359.  Dave Bacon is also supported by ARO/NSA Quantum Algorithms Grant No. W911NSF-06-1-0379 and NSF Grant No. 0621621.

Some work was done during the summer of 2007, during which Gregory Crosswhite was supported by the East Asia Pacific Summer Institute (EAPSI) Program, which was cofunded by the NSF and the Australian Academy of Science.
}


\appendix


\section{Visualization of matrix degree of freedom}

\begin{figure}
\subfloat[Matrix product diagram for ``cat'' state.
   \label{fig:cat-mps}
]{\framebox{\includegraphics[width=\columnwidth]{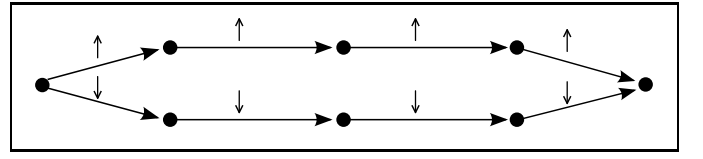}}}
\subfloat[Nodes are rearranged in the middle of the diagram.
   \label{fig:cat-mps-with-twist}
]{\framebox{\includegraphics[width=\columnwidth]{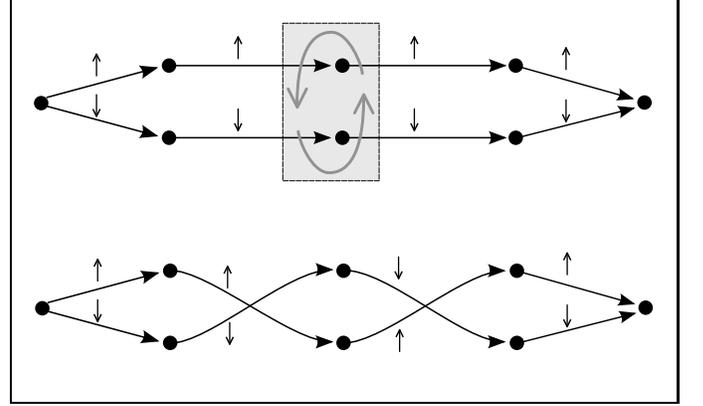}}}
\subfloat[A connection is added which results in two additional walks that cancel each other.
   \label{fig:cat-mps-extraneous-paths}
]{\framebox{\includegraphics[width=\columnwidth]{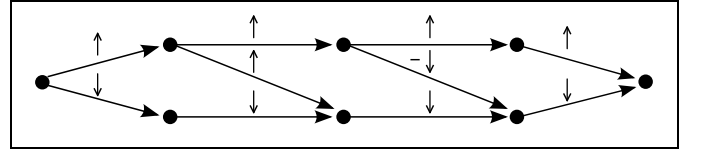}}}
\subfloat[A node is inserted in the middle of the diagram.
   \label{fig:cat-mps-node-inserted}
]{\framebox{\includegraphics[width=\columnwidth]{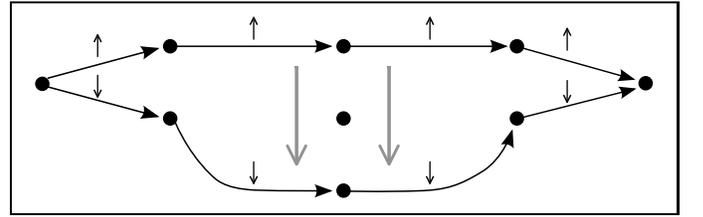}}}
\caption{The matrix product diagram is given for a 4-particle ``cat'' state.  Some transformations are demonstrated that result in equivalent representations for this state.}
\end{figure}

There is a matrix ``degree of freedom'' that allows us to manipulate our state into equivalent forms.  To visualize this process, we consider the 4-particle ``cat'' state,

$$
\uparrow\uparrow\uparrow\uparrow + \downarrow\downarrow\downarrow\downarrow =
\begin{bmatrix}\uparrow & \downarrow \end{bmatrix}\cdot
\bmat{\uparrow}{0}{0}{\downarrow}\cdot
\bmat{\uparrow}{0}{0}{\downarrow}\cdot 
\begin{bmatrix}\uparrow \\ \downarrow \end{bmatrix},
$$

\noindent with the corresponding diagram shown in \ref{fig:cat-mps}.

The matrix degree of freedom comes from the fact that we may insert any matrix $X$ and its inverse $X^{-1}$ into this product without altering it:

$$
\begin{aligned}
  \Psi &= \underbrace{\begin{bmatrix}\uparrow & \downarrow \end{bmatrix}}_{A}\cdot
         \underbrace{\bmat{\uparrow}{0}{0}{\downarrow}}_{B}\cdot X \cdot X^{-1}
         \underbrace{\bmat{\uparrow}{0}{0}{\downarrow}}_{C}\cdot 
         \underbrace{\begin{bmatrix}\uparrow \\ \downarrow \end{bmatrix}}_{D}\\
       &= \underbrace{\begin{bmatrix}\uparrow & \downarrow \end{bmatrix}}_{A}\cdot
         \underbrace{\paren{\bmat{\uparrow}{0}{0}{\downarrow}\cdot X}}_{B'} \cdot 
         \underbrace{\paren{X^{-1}\cdot\bmat{\uparrow}{0}{0}{\downarrow}}}_{C'}\cdot
         \underbrace{\begin{bmatrix}\uparrow \\ \downarrow \end{bmatrix}}_{D}
\end{aligned}
$$

This degree of freedom gives us many ways to alter our diagram to produce equivalent representations.  For example, we may rearrange the nodes at any point in the diagram, as shown in figure \ref{fig:cat-mps-with-twist}, which is equivalent to introducing the matrix

$$X = \bmat{0}{1}{1}{0}\quad\Rightarrow\quad \Psi =\begin{bmatrix}\uparrow & \downarrow \end{bmatrix}\cdot
         \bmat{0}{\uparrow}{\downarrow}{0}\cdot
         \bmat{0}{\uparrow}{\downarrow}{0}\cdot 
         \begin{bmatrix}\downarrow \\ \uparrow \end{bmatrix} $$

We may also add additional connections between nodes in our diagrams, as long as the additional paths which result cancel each other out.  So for example, we may modify our diagram to become as shown in \ref{fig:cat-mps-extraneous-paths}.  Note how the paths introduces by the two new connections have opposite phases, and thus cancel.  This change corresponded to introducing the matrix

$$X = \bmat{1}{1}{0}{1} \quad \Rightarrow\quad
  \Psi =\begin{bmatrix}\uparrow & \downarrow \end{bmatrix}\cdot
         \bmat{\uparrow}{\uparrow}{0}{\downarrow}\cdot
         \bmat{\uparrow}{-\downarrow}{0}{\downarrow}\cdot 
         \begin{bmatrix}\downarrow \\ \uparrow \end{bmatrix}.
$$

$X$ need not be a square matrix, as long as the product $X \cdot X^{-1}$ results in an identity matrix with the appropriate dimensions.  This gives us the freedom to introduce empty nodes into our diagram, moving the connected noted apart, as with the matrix

$$X = \begin{bmatrix}
1 & 0 & 0\\
0 & 0 & 1\\
\end{bmatrix},$$

\noindent which results in the diagram shown in \ref{fig:cat-mps-node-inserted}.

The fact that we moved them apart means that we should be able to move them back together again; this can be done with the matrix,

$$X = \begin{bmatrix}
1 & 0\\
0 & 0\\
0 & 1\\
\end{bmatrix}.$$

This matrix does not have the property that $X \cdot X^{-1}$ gives us an identity; rather, it gives us a projector onto a two-dimensional subspace.  This is perfectly fine, however, as in this case the third dimension (corresponding to the unused node) was actually redundant.  It is obvious for this diagram that this was the case, but it might not be so obvious for a general diagram.  Nonetheless, numerically it is easy to eliminate such redundant nodes between two matrices by contracting the two tensors along their adjoining index and then using a singular value decomposition to split them back apart, dropping all (post-decomposition) vertices which correpond to zero singular values.

\section{Periodic boundary conditions}

Up to this point we have been implicitly assuming open boundary conditions on our system.  In many situations, however, one wants to impose periodic boundary conditions.  To do this, an additional bond is added between the first and last matrix:

$$\Psi^{\alpha\beta\gamma\delta\dots} = \sum_{i,j,k,\textbf{z}} A_{\textbf{z}i}^\alpha B_{ij}^\beta C_{jk}^\gamma D_{k\textbf{z}}^\delta.$$

Our automata interpretation is modified as follows:  rather than having initial and final distributions (i.e., starting states and ending states), we instead allow the automata to start on any state, but restrict it to only accept strings which cause it to end on the state at which it started.  This is equivalent to revising \eqref{wfa-compute-eqn} to replace the initial distribution $\alpha$ and the final distribution $\Omega$ with a trace,

$$f(a_0a_1\dots a_N) = \tr\paren{ W_{a_0} \cdot W_{a_1} \dots W_{a_N} }.$$

This idea generalizes straightforwardly to multiple dimensions.

\bibliography{caching}

\end{document}